\documentclass[a4paper,11pt]{article}

\pdfoutput=1 

\usepackage[T1]{fontenc} 

\usepackage[]{caption}
\usepackage{color,xcolor,graphicx,amsmath, amssymb,mathtools,hyperref
, physics,ulem}

\title{{\bf Balancing asymmetric dark matter with baryon asymmetry and dilution of frozen dark matter by sphaleron transition}}
\author{Arnab~Chaudhuri$^{a}$\footnote{{\bf e-mail}: \href{mailto:arnabchaudhuri.7@gmail.com}{arnabchaudhuri.7@gmail.com}},
Maxim Yu. Khlopov$^{b}$\footnote{{\bf e-mail:} \href{mailto:khlopov@apc.in2p3.fr}{khlopov@apc.in2p3.fr}},
\\
$^a$ \small{ Novosibirsk State University} \\
\small{ Pirogova ul., 2, 630090 Novosibirsk, Russia}\\
$^b$ \small{Institute of Physics, Southern Federal University}\\
\small{Stachki 194 Rostov on Don 344090, Russia}\\
\small{and Université de Paris, CNRS, Astroparticule et Cosmologie, F-75013 Paris, France}\\
\small{and National Research Nuclear University "MEPHI", Moscow 115409, Russia}\\
}

\date{\today}

\begin{document}

\maketitle

\begin{abstract}
In this paper we study the effect of electroweak sphaleron transition or electroweak phase transition (EWPT) in balancing baryon excess to the excess stable quarks of $4^{th}$ generation. Considering the conservation of all quantum number and charges, sphaleron transition between between baryons, leptons and $4^{th}$ family of leptons and quarks is possible. We have tried to established a possible definite relationship between the value and sign of the $4^{th}$ family excess relative to baryon asymmetry under the framework of second order EWPT. In passing by we show the small, yet negligible dilution in the pre-existing dark matter density due the sphaleron transition.
\end{abstract}

\section{Introduction}
The origin of the baryon asymmetry of the universe (BAU) has been dealt with great interest and passion throughout the last few decades, see~\cite{RevModPhys.53.1}, \cite{cohen} and \cite{cohen2}. Numerous models of BAU are made, ranging from
BAU generation at the GUT stage of the Universe to the generation
during the electroweak phase transition (EWPT). Irrespective of the the mechanisms, the preexisting
asymmetry is diluted by baryon number violating mechanisms in the
electroweak theory. This is due to the violation of the baryon and lepton number and the non-trivial topological structure
of the Yang–Mills theory. Topologically speaking individual vacua are expressed by Chern–Simons number $N_{CS}$, see~\cite{PhysRevLett.37.172} and they are located at barrier of energy of the magnitude of $m_W/\alpha$. Here $m_W$ is the mass of $W$-boson and $\alpha$ is the $SU(2)$ gauge coupling constant and is given as $\alpha=g^2/(4\pi)$.

In electroweak theory, it is possible for a sphaleron to convert baryons into antileptons and vice versa, thus violating the baryon (lepton) number. And in this process the difference between the baryon and lepton numbers $(B-L)$ is conserved, even though individually the quantum numbers are violated. Hence it is important to know about the transition rates of such process.

Sphalerons are also associated with saddle points \cite{KM}. It is interpreted as the peak energy configuration. It can be demonstrated that because of this anomaly transition between vacua are associated with a violation of Baryon (and lepton) number.

Although baryon excess can be created at the time of electroweak symmetry breaking, it is preserved during first order phase transition. In second order, sphalerons can wipe out the total asymmetry created but in first order, only the asymmetry created in the unbroken phase is wiped out.

There are many beyond standard model theories which resurfaced in the recent years related to the cosmological dark matter, the most recent one being channels from the $g-2$ muon  \cite{Abi:2021gix,Arcadi:2021yyr,Chen:2021jok}. However, the cosmological dark matter may have dark atoms, where stable charged particles of new origin are bounded by Coulomb interaction. Many multi-charged models can exists where $\pm 2$ charged stable species are seen. Firstly, there are the AC-leptons which uses approximate cummutative geometry see~\cite{MYK,SAC}. Secondly, the walking technicolor (WTC) model, based on techni-baryons, see~\cite{12,13,14,15,16,17,18,19} and leptopns and finally the $(\bar{U}\bar{U}\bar{U})$, can be named as heavy quark clusters, consisting of anti-$U$ quark of the $4^{th}$ generation, see~\cite{20,21,22,23}. These extra charges combined with electrons in helium can be restricted to ordinary terrestrial matter. 

Multiple cosmological scenarios can provide methods for suppressing this anomalous helium. Firstly, $U(1)$ gauge symmetry causing Coulomb like interactions between charged dark matter particles and thus masking the existence of hidden photons. Secondly, in the early universe, the excess of anomalous helium is negligible due to the suppression of free charges. 

In this work we consider $4^{th}$ generation family as an extension to the Standard Model (SM) and proceed to study the electroweak phase transition (EWPT). The simplest charge-neutral model is considered here and also we've considered that EWPT is of second order. In passing by, we showed the dilution of pre-existing frozen out dark matter density in the presence of the $4^{th}$ generation. 

The paper is organised as follows: In the next section we talk about $4^{th}$ generation family, defining definite relationship between the value and sign of $4^{th}$ generation family excess relative to the baryon asymmetry due to electroweak phase transition and possible sphaleron production has been established. It is followed by calculation of the dilution factor of pre-existing dark matter density and it follows by a general conclusion.

\section{$4^{th}$ generation: A brief Review and linking baryon excess to excess of stable $4^{th}$ generation fermions}

The family of elementary particles consists of fermions and gauge bosons based on the spin: $1/2$ for fermions and integer for bosons. Fermions are further divided into two more two more families of quarks and leptons, which are arranged into three generations with two leptons and two quarks in each generation. Leptons consists of electron like ($-1$ charged) particles and neutrino like ($0$ charged) particles. The quarks fall in the category of down type ($-1/3$ charge) and up type ($+2/3$ charge). The particles interact with the exchange of gauge bosons. The fundamental fermions and their interactions by exchange of intermediate bosons are described by the Standard Model of particle physics.  The SM predicts also the existence of the Higgs boson to be discovered, which is responsible for the generation of particles' masses within the Standard Model.

All the mentioned particles have been experimentally verified. The fourth generation is of theoretical interest in the context of sphaleron transition, electroweak symmetry breaking and large CP violating processes in the $4 \times 4$ CKM matrix which might play a crucial role in understanding the baryon asymmetry in the Universe. Thus, there are significant efforts ongoing to search for the fourth generation. 

Due to the excess of $\bar{{U}}$, only $-2$ charge or neutral hadrons are present in the universe. $^4He$ formed during Big Bang nucleosynthesis completely screens $Q^{--}$ charged hadrons in composite $[^4HeQ^{--}]$ "atoms". If this $4^
{th}$ family follows from string phenomenology, we have new charge associated ($F$) with $4^{th}$ family fermions. Principally $F$ should be the only conserved quantity but to keep matter simple, an analogy with WTC model is made and assume two numbers $FB$ (for 4th quark) and $L'$ for ($4^{th}$ neutrino). Detailed calculations of WTC have been done in
 \cite{12,24} and most of the terminology are kept same as the above mentioned papers.

Most of the asymmetries generated before EWPT can be destroyed by quantum anomalies. Individual quantum numbers of individual species might not be conserved but their differences are conserved. These process, otherwise known as "Sphaleron" processes, might be negligible in the present days but they were active during the time the Universe had a temperature above or at the scale of the electroweak symmetry breaking. As universe expanded and the temperature fell down and the quantum number violating processes ceases to exists, the relation among the particles emerging from the process (SM+$4^{th}$ generation) follows the expression:
\begin{equation} \label{5}
   3(\mu_{u_L}+\mu_{d_L})+\mu+\mu_{U_L}+\mu_{D_L}+\mu_{L'}=0. 
\end{equation}
Here $\mu$ is the chemical potential of all the SM particles, $\mu_{L'}$ is the chemical potential of the new species leptons and $\mu_{U_L}$ and $\mu_{D_L}$ are that of the $4^{th}$ generation quarks, see~\cite{24}. The number densities follows, respectively for bosons and feremions:
\begin{equation} \label{6}
    n=g_*T^3\frac{\mu}{T}f(\frac{m}{T}),
\end{equation}
and
\begin{equation} \label{7}
    n=g_*T^3\frac{\mu}{T}g(\frac{m}{T}),
\end{equation}
where $f$ and $g$ are hyperbolic mathematical functions and $g_*$ is the effective degrees of freedom and they are given by:
\begin{equation} \label{7a}
    f(z)=\frac{1}{4\pi^2}\int_0^{\infty}x^2 cosh^{-2}\left(\frac{1}{2}\sqrt{z^2+x^2} \right)dx,
\end{equation}
and
\begin{equation} \label{7b}
    g(z)=\frac{1}{4\pi^2}\int_0^{\infty} x^2 sinh^{-2}\left(\frac{1}{2}\sqrt{z^2+x^2} \right)dx.
\end{equation}

The number density of baryons follows the expression:
\begin{equation} \label{8}
    B=\frac{n_B-n_{\bar{B}}}{gT^2/6}.
\end{equation}
As the main point of interest is the ratio of baryon excess to to excess of stable $4^{th}$ generation, the normalization cancels out without without loss of generality. 

Let us define a parameter $\sigma$ such that, which respectively for fermions and bosons are given by:
\begin{equation} \label{9}
    \sigma=6f\frac{m}{T_c},
\end{equation}
and
\begin{equation} \label{10}
    \sigma=6g\frac{m}{T_c}.
\end{equation}
$T_c$ is the transition temperature and is given by
\begin{equation} \label{10a}
    T_c=\frac{2M_W(T_c)}{\alpha_W~ln(M_{pl}/T_c)}B(\frac{\lambda}{\alpha_W}).
\end{equation}
In the above equation $M_W$ is the mass of W-boson, $M_{pl}$ is the Planck mass and $\lambda$ is the self-coupling. The function $B$ is derived from experiment and takes the value from $1.5-2.7$.

The new generation charge is calculated to be:
\begin{equation} \label{11}
    FB=\frac{2}{3}(\sigma_{U_L}\mu_{U_L}+\sigma_{U_L}\mu_{D_L}+\sigma_{D_L}\mu_{D_L}),
\end{equation}
where $FB$ corresponds to the anti-U ($\bar{U}$) excess. For detailed calculation please see \textbf{cite 24}.

The SM baryonic and leptonic quantum numbers are expressed as:
\begin{equation} \label{10a1}
    B=\left[(2+\sigma_t)(\mu_{uL}+\mu_{uR})+3(\mu_{dL}+\mu_{dR}) \right]
\end{equation}
and
\begin{equation} \label{10b}
    L=4\mu+6\mu_W
\end{equation}
where in eqn. \ref{10a1} the factor 3 of the down-type
quarks is the number of families. For the $4^{th}$ generation lepton family, quantum number is given by:
\begin{equation} \label{10c}
    L'=2(\sigma_{\nu'}+\sigma_{U_L})\mu_{\nu'L}+2\sigma_{U_L}\mu_W+(\sigma_{\nu'}-\sigma_{U_L})\mu_0
\end{equation}
where $\nu'$ is the new family of neutrinos originating from the extension of SM and $\mu_0$ is the chemical potential from the Higgs sector in SM.

Due to the presence of a single Higgs particle, the phase transition is of second order. The ratio of the number densities of $4^{th}$ generation to the baryons is determined by:
\begin{equation} \label{12}
    \frac{\Omega_{FB}}{\Omega_B}=\frac{3}{2}\frac{FB}{B}\frac{m_{FB}}{m_p}.
\end{equation}
Electrical neutrality and negligibly small chemical potential of Higgs sector is the result of second order phase transition. The ratio of the number density of number density of the $4^{th}$ generation to the baryon number density can be expressed as a function of the ratio of original and new quantum numbers. In the limiting case of second order EWPT, we get:
\begin{equation} \label{13}
    -\frac{FB}{B}=\frac{\sigma_{U_L}}{3(18+\sigma_{\nu'})}\left[(17+\sigma_{\nu'})+\frac{(21+\sigma_{\nu'})}{3}\frac{L}{B}+\frac{2}{3}\frac{9+5\sigma_{\nu'}}{\sigma_{\nu'}}\frac{L'}{B}\right].
\end{equation}
Hence we establish a relationship between the baryon excess and the excess of $\bar{U}$ for second order EWPT.

\section{Dilution of pre-existing dark matter density}
The thermodynamic quantity, entropy density, is a conserved quantity in the initial stage of universe expansion, especially when the primeval plasma was in thermal equilibrium with negligible chemical potential. As soon as the universe goes into the state of thermal non-equilibrium, i.e., when $\Gamma>H$, where $\Gamma$ is the reaction rate and $H$ is the Hubble parameter, the conservation law breaks down and entropy starts pouring into the plasma and this can dilute the pre-existing baryon asymmetry and dark matter density. 

There are many instances of entropy production, like primordial black hole evaporation~\cite{Chaudhuri:2020wjo}, Electroweak phase transition within standard model and two Higgs doublet model~\cite{Chaudhuri:2017icn},\cite{Chaudhuri:2021agl} and \cite{AMS}. Apart from these, the freeze out of dark matter density might lead to entropy production, which in turn can dilute the pre-existing dark matter density. 

The Lagrangian of theory consists of the Langrangian of the standard model (SM) and the interaction terms of the $4^{th}$ generation fermionic family. It is given by:
\begin{equation} \label{lag}
    \mathcal{L}=\mathcal{L}_{SM}+\mathcal{L}_{4^{th}}, 
\end{equation}
where $\mathcal{L}_{SM}$ is given by
\begin{equation}
    \mathcal{L}_{SM}=\frac{1}{2}g^{\mu \nu}\partial_{\mu}\phi \partial_{\nu}\phi -U_{\phi}(\phi)+\sum_{j} i\left[g^{\mu \nu} \partial_{\mu} \chi_j^{\dagger} \partial_{\nu} \chi_j - U_j(\chi_j) \right].
\end{equation}
The CP violating potential of the theory follows as:
\begin{equation}
    U_{\phi}(\phi)=\frac{\lambda}{4}\left(\phi^2-\eta^2 \right)^2+\frac{T^2 \phi^2}{2}\sum_j h_j \left(\frac{m_j(T)}{T} \right).
\end{equation}
Here $\lambda=0.13$ is the quartic coupling constant and $\eta$ is the vacuum expectation values which is $\sim 246~GeV$ in the SM. $T$ is the plasma temperature and $m_j(T)$ is the mass of the $\chi_j$-particle at temperature T, see~\cite{imelo}.

To calculate the dilution factor, it is necessary to compute the energy and the pressure density of the plasma using the energy-momentum tensor. Following the detailed calculation in ~\cite{Chaudhuri:2021agl} and assuming the universe was flat in the early epoch with the metric $g_{\mu \nu}=(+,-,-,-)$ we have
\begin{equation} \label{p+r}
    \rho+\mathcal{P}=\dot{\phi}^2+\frac{4}{3}\frac{\pi^2 g_*}{30}T^4.
\end{equation}

The last term in Equation (\ref{p+r}) arises from the Yukawa interaction between fermions. The Higgs field starts to oscillate around the minimum which appears during the phase transition. Particle production from this oscillating field causes the damping. The characteristic time of decay is equal to the decay width of the Higgs bosons. If it is large in compared to 
the expansion and thus the universe cooling rate, then we may assume that Higgs bosons essentially live in the minimum of the potential. This has being clearly discussed in~\cite{Chaudhuri:2017icn}.

To calculate the entropy production, it is necessary to solve the evolution equation for energy density conservation, 
\begin{equation} \label{fried}
\Dot{\rho}=-3\mathcal{H}(\rho+P).
\end{equation}

\begin{figure}[h!]
\centering
\includegraphics[]{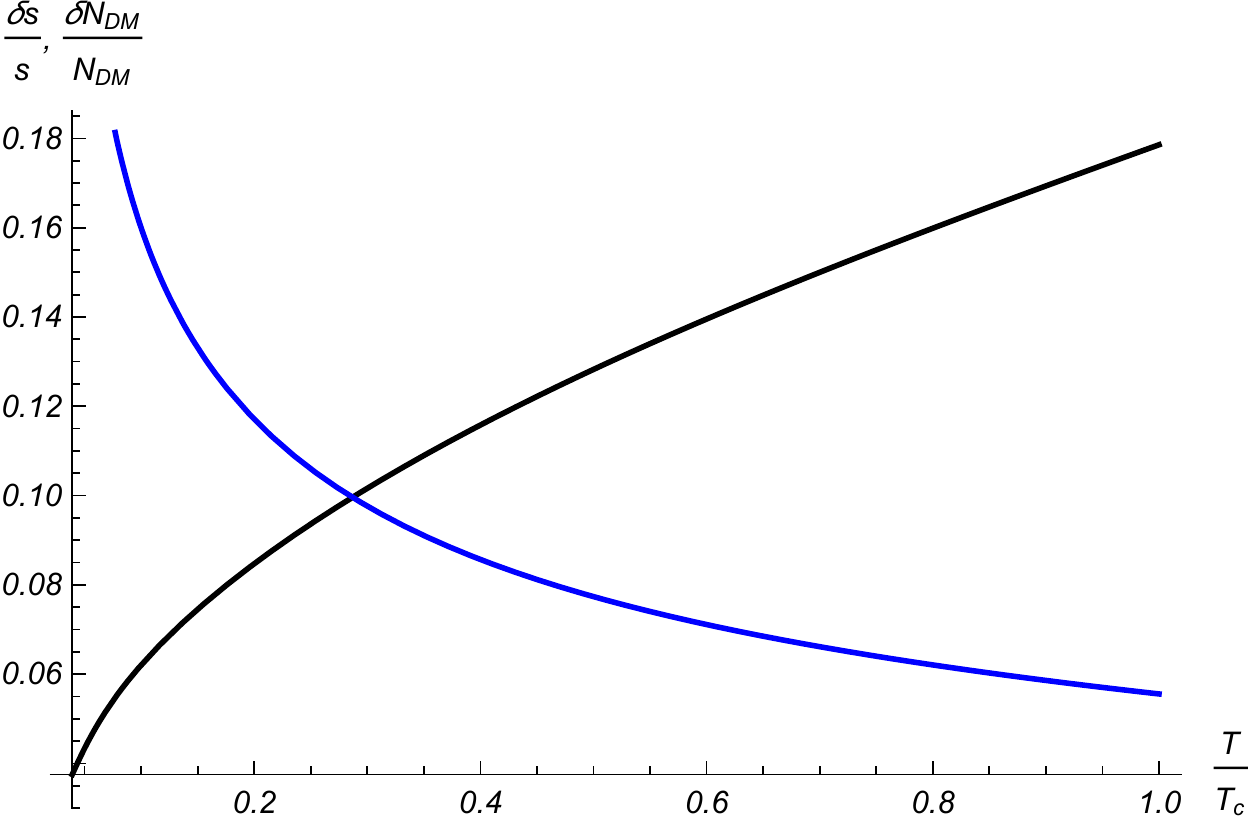}
\caption{Entropy production (black line) and the dilution of pre-existing dark matter (blue line) in the presence of $4^{th}$ generation fermions are presented.  }
\label{f-entropy-1}
\end{figure}

In figure (\ref{f-entropy-1}), both the dilution of the pre-existing dark matter (blue line) and the entropy production in the presence of $4^{th}$ generation lepton family (black line) are shown. It is clear that since the sphaleron transition is of second order, the net dilution and entropy production $(\sim 18\%)$ is somewhat low compared to the scenarios of first order. Again, the presence of a single Higgs field made the phase transition second order.

\section{Conclusions}

It is clear from eqn. \ref{13} that a definite relationship between the value and sign of $4^{th}$ generation family excess relative to the baryon asymmetry due to electroweak phase transition and possible sphaleron production has been established. Dark matter candidates in the form of bounded dark atom can emerge from this model due to the excess of $\bar{U}$ within primordial He nuclei. We have considered only the lightest and most stable particles and also took into account only the second order phase transition. Dilution of pre-existing dark matter density is calculated and in the present scenarios the dark matter density is reduced by $\sim 18\%$. The situation can change more drastically if unstable particle and first order phase transition were taken into account but those scenarios are beyond the scope of this paper. 




\section{Author Contribution}
Article by A.C. and M.K. The authors contributed equally to this~work.
All authors have read and agreed to the published version of the manuscript.

\section{Funding}
The work of A.C. is funded by RSF Grant 19-42-02004. The research by M.K.was financially supported by grant of the Russian Science Foundation (Project No-18-12-00213-P).

\section{Conflict of Interest}
There has been no conflict of interest among the authors of this paper

\end{document}